# Rankings of countries based on rankings of universities

Bahram Kalhor[1], Farzaneh Mehrparvar[2]

## Abstract

Although many methods have been designed for ranking universities, there is no suitable system that focuses on the ranking of countries based on the performance of their universities. The overall ranking of the universities in a region can indicate the growth of interests in science among the people of that land. This paper introduces a novel ranking mechanism based on the rankings of universities. Firstly, we introduce and discuss two new rankings of countries, based on the rank of their universities. Secondly, we create rankings of countries according to the selected method, based on the top 12000 universities in webometrics.info (January 2012) and compare rankings of countries in 4 editions (January 2012 to July 2013). Firstly, we introduce two new methods of ranking countries based on their university rankings, Weighted Ranking (WR) and Average Ranking (AR). Secondly, we discuss how the introduced ranking systems, perform in ranking countries based on the two years of data. Thirdly, we choose QS (http://www.topuniversities.com) and webometrics.info as two different classification systems for comparing rankings of countries, based on the top 500 universities in these rankings. Results indicate that the methodology can be used to show the quality of the whole universities of each country used to compare rankings of countries in practice compare to other countries in the world.

## Introduction

Unlike rankings of universities, rankings of countries have not yet been the focus of considerable study in the web rankings research. There are many countries in which most of their universities belong to the government. In these countries, universities must follow their country policies, especially on the internet. Finding rankings of countries based on rankings of universities help researchers for comparing the performance of government decisions in the educational institutes and universities.

Higher education institutions are using rankings as a promotion tool that shows their educational and research excellence. Universities use these rankings for increasing their research performance (Isidro F. Aguillo, Judit Bar-lan, Mark Levene, Jose Luis Ortega, 2010). Local and international rankings have been focused by higher education policymakers (Ghane, Khosrowjerdi & Azizkhani, 2013). Rankings of the Times Higher Education (THE-QS, http://www.topuniversities.com/home), rankings of the Shanghai Jiao Tong University (ARWU, http://www.arwu.org), rankings of the Higher Education and Accreditation Council of Taiwan (HEEACT, http://ranking.heeact.edu.tw), web rankings of World Universities by the Cybermetrics Lab at CSIC (WR, http://www.webometrics.info), web rankings of Iranian Universities (RICEST, http://websanji.ricest.ac.ir), rankings of the Centre for Science and Technology Studies at Leiden University (CWTS, http://ranking.heeact.edu.tw), rankings of the SCImago Research Group (SJR, http://www.scimagoir.com) and 4 International Colleges & Universities (http://www.4icu.org) are the famous rankings of universities.

---

[1] Department of Computer, College of Mechatronic, Karaj Branch, Islamic Azad University, Alborz, IRAN
[2] Department of Physics, Karaj Branch, Islamic Azad University, Alborz, IRAN

The concept of using rankings of universities for calculating countries' educational rankings is the central focus of this article. We used the data from the 4 data set updates, in two years, on university rankings of Webometrics. We produce country rankings from the January 2012 list and compare their rankings to the July 2012, January 2013, and July 2013. WR is a web-based ranking (21451 universities, January 2014) which is published every six months. We use a list of 12000 universities from each edition. We study and discuss changes in countries' ranks in one ranking of universities (WR) among the time. We also compare the rankings of countries in two different classification systems. We compare the rank of countries in the top 500 WR (January 2012) with the countries' rank in the top 500 THE-QS (2012).

**Methodology**

For quantitative purposes, data of universities and research centers have been extracted from webometrics.info and QS. Although webometrics.info only provides the latest rankings of universities, in the last three years we have saved a full list of rankings of universities from Webometrics.info. Webometrics.info collects data for all the universities of the world. The latest main list webometrics.info contains the top 12000 domains (http://webometrics.info/en/world?page=119). Webometrics' previous editions have only the top 500 universities from July 2009 (http://webometrics.info/en/Previous_editions). Many countries don't have any university in the top 500 (http://webometrics.info/en/node/54) thus in the last three years we have saved a full list of rankings of universities (top 12000 domains) from Webometrics.info in our database. In this article, we use this database for calculating rankings of countries. In this study, we also produce QS countries' rank based on the top 500 QS World University Rankings (2012, http://www.topuniversities.com/university-rankings/world-university-rankings/2012) and compare with WR countries' rank (webometrics.info) based on top 500 WR universities (January 2012). To compare the rankings of countries in each edition top 50 world countries with the highest number of universities in the WR (January 2012) were selected (Table 1). In this list United States of America with 2883 universities (January 2012) is at the top and Egypt with 34 universities is at the bottom of the list. Selected top 50 countries list contains 10927 universities (Table 1).

In the end, we used the top 500 QS universities (QS, 2012) for comparing to the top 500 WR (WR, January 2012) universities. Although Webometrics.info provides 12000 universities in each edition and QS provides 750 universities, we used the top 500 universities. The simplest way of calculating the rank of one country is by calculating average rank. We define the average rank for one country as the sum of the universities rank belongs to that country divide to count of universities.

$$AR = \frac{\sum_{i=1}^{n} R_i}{n}$$

Where AR is the Average Rank of the country and n is the total number of universities that belong to the country. Country average rank (AR), has two major problems drawback. The first problem is that items of each edition are not identical and comparing overlapping universities is not recommended. In this case, exiting one university at the bottom of the rankings list makes better AR for the country. The Second problem issue is the huge differences between the number of universities in each country. One country with a lower number of universities maybe takes a better position in the ranking table rather than a country like the United States of America which occupied the most position of top 100 universities (Table 4). To address the drawbacks of the AR problems system, we defined W as a new indicator that allows us to detect the weight of countries.

$$W = \sum_{i=1}^{n} (M - R_i + 1)$$

Where the W is the weight of the country, n is the total number of universities number of each country, M is the total number of universities number in the world's list (12000 in this study) and i is the ranking of each university in the Webometrics.info. W has the value zero when the country does not have any university at the world's list, its maximum value is M*(M+1)/2. Table 6 shows the calculated W according to the universities ranking for the top 50 countries in WR for 4 editions of different data sets. The Maximum W in July 2012 belongs to the United State of America with 18943854 and the lowest W belongs to Morocco with 103988.

M=12000

Min W=0

Max W= M*(M+1)/2 =7200600

1+2+3+4+…. + M=M*(M+1)/2

Both problems of AR are solved by using W.

**Results**

Weighted ranking (W) provides a new way to compare countries. We can compare changes of countries' rankings in this new ranking's list to other official rankings' list. Also, we can compare two separate rankings' lists. In the top 500 QS, there are 50 countries (Table 2) and in the top 500 WR, there are 48 countries (Table 3).

## Comparing four editions of WR

Table (6) shows the ranks of 50 countries in four WR editions. Selected countries are the countries with the highest universities number in January 2012, ranks have been produced based on country weight (table 5). The United State of America, china, and japan are at the top of the list. The highest drop from January 2012 to July 2013 belongs to Romania with 9 levels and the highest increase belongs to Pakistan with 5 levels.

Table 1: Top 50 countries with highest number of universities in top 12000 WR list

| Rank | Country Name | January 2012 | July 2012 | January 2013 | July 2013 | Rank | Country Name | January 2012 | July 2012 | January 2013 | July 2013 |
|---|---|---|---|---|---|---|---|---|---|---|---|
| 1 | United States of America | 2883 | 2750 | 2829 | 3031 | 26 | Vietnam | 85 | 82 | 83 | 77 |
| 2 | China | 1114 | 1073 | 1117 | 1102 | 27 | Romania | 85 | 76 | 78 | 67 |
| 3 | Japan | 679 | 645 | 758 | 773 | 28 | Switzerland | 84 | 73 | 70 | 81 |
| 4 | Brazil | 576 | 490 | 514 | 366 | 29 | Australia | 72 | 63 | 70 | 78 |
| 5 | Russian Federation | 447 | 418 | 640 | 577 | 30 | Philippines | 72 | 68 | 64 | 61 |
| 6 | India | 378 | 305 | 382 | 478 | 31 | Portugal | 70 | 66 | 69 | 64 |
| 7 | France | 372 | 337 | 397 | 394 | 32 | Austria | 66 | 64 | 64 | 64 |
| 8 | Germany | 337 | 317 | 324 | 341 | 33 | Belgium | 64 | 56 | 61 | 58 |
| 9 | Republic Of Korea | 286 | 284 | 272 | 245 | 34 | Chile | 64 | 60 | 56 | 57 |
| 10 | Poland | 265 | 236 | 273 | 270 | 35 | Malaysia | 64 | 62 | 60 | 64 |
| 11 | United Kingdom | 218 | 203 | 243 | 257 | 36 | Peru | 63 | 64 | 64 | 54 |
| 12 | Mexico | 201 | 182 | 153 | 140 | 37 | Czech Republic | 61 | 56 | 51 | 54 |
| 13 | Iran (Islamic Republic of Iran) | 178 | 163 | 161 | 172 | 38 | Hungary | 57 | 51 | 52 | 48 |
| 14 | Indonesia | 162 | 161 | 188 | 160 | 39 | Denmark | 56 | 49 | 47 | 45 |
| 15 | Taiwan | 154 | 153 | 151 | 154 | 40 | Greece | 51 | 47 | 49 | 52 |
| 16 | Canada | 154 | 148 | 171 | 231 | 41 | Norway | 51 | 48 | 49 | 47 |
| 17 | Spain | 154 | 139 | 137 | 131 | 42 | Finland | 48 | 47 | 47 | 45 |
| 18 | Italy | 141 | 123 | 126 | 132 | 43 | Ecuador | 44 | 40 | 37 | 30 |
| 19 | Thailand | 140 | 135 | 135 | 130 | 44 | Kazakstan | 44 | 37 | 40 | 41 |
| 20 | Colombia | 139 | 134 | 133 | 95 | 45 | Sweden | 43 | 42 | 43 | 42 |
| 21 | Turkey | 131 | 129 | 140 | 140 | 46 | Ireland | 41 | 34 | 31 | 36 |

| Rank | Country Name | January 2012 | July 2012 | January 2013 | July 2013 | Rank | Country Name | January 2012 | July 2012 | January 2013 | July 2013 |
|---|---|---|---|---|---|---|---|---|---|---|---|
| 22 | Ukraine | 112 | 106 | 184 | 126 | 47 | Bulgaria | 40 | 41 | 40 | 40 |
| 23 | Argentina | 98 | 92 | 96 | 98 | 48 | Venezuela | 37 | 35 | 32 | 29 |
| 24 | Netherlands | 90 | 76 | 78 | 71 | 49 | Morocco | 36 | 28 | 29 | 36 |
| 25 | Pakistan | 86 | 72 | 74 | 85 | 50 | Egypt | 34 | 34 | 37 | 34 |

Table 2: Top countries with highest number of universities at the top 500 QS (2012)

| Rank | Country Name | University | Rank | Country Name | University |
|---|---|---|---|---|---|
| 1 | United States of America | 99 | 26 | Austria | 5 |
| 2 | United Kingdom | 51 | 27 | Brazil | 5 |
| 3 | Germany | 39 | 28 | Colombia | 4 |
| 4 | Australia | 24 | 29 | Saudi Arabia | 4 |
| 5 | France | 22 | 30 | Palestine | 4 |
| 6 | Canada | 20 | 31 | Norway | 4 |
| 7 | Japan | 20 | 32 | Chile | 3 |
| 8 | China | 18 | 33 | Portugal | 3 |
| 9 | Spain | 14 | 34 | Indonesia | 3 |
| 10 | Italy | 14 | 35 | South Africa | 3 |
| 11 | Netherlands | 13 | 36 | Greece | 2 |
| 12 | Republic Of Korea | 13 | 37 | Kazakstan | 2 |
| 13 | Taiwan | 11 | 38 | Mexico | 2 |
| 14 | Finland | 8 | 39 | Philippines | 2 |
| 15 | Switzerland | 8 | 40 | Poland | 2 |
| 16 | Sweden | 8 | 41 | Singapore | 2 |
| 17 | India | 7 | 42 | Thailand | 2 |
| 18 | Ireland | 7 | 43 | Turkey | 2 |
| 19 | Belgium | 7 | 44 | United Arab Emirates | 2 |
| 20 | New Zealand | 7 | 45 | Uruguay | 1 |
| 21 | Malaysia | 6 | 46 | Egypt | 1 |
| 22 | Russian Federation | 6 | 47 | Czech Republic | 1 |
| 23 | Hong Kong | 6 | 48 | Oman | 1 |
| 24 | Denmark | 5 | 49 | Pakistan | 1 |
| 25 | Argentina | 5 | 50 | Lebanon | 1 |

Table 3: Top countries with highest number of universities at the top 500 WR (January 2012)

| Rank | Country Name | University | Rank | Country Name | University |
|---|---|---|---|---|---|
| 1 | United States of America | 155 | 25 | Greece | 4 |
| 2 | Germany | 44 | 26 | Ireland | 4 |
| 3 | United Kingdom | 30 | 27 | Palestine | 4 |
| 4 | Spain | 24 | 28 | Turkey | 4 |
| 5 | Canada | 23 | 29 | Norway | 4 |
| 6 | Australia | 17 | 30 | Czech Republic | 3 |
| 7 | Italy | 16 | 31 | Indonesia | 3 |
| 8 | Taiwan | 14 | 32 | South Africa | 3 |
| 9 | China | 14 | 33 | Hungary | 3 |
| 10 | Japan | 12 | 34 | New Zealand | 3 |
| 11 | Brazil | 12 | 35 | Malaysia | 3 |
| 12 | Netherlands | 11 | 36 | Argentina | 2 |
| 13 | Sweden | 9 | 37 | Slovakia | 2 |
| 14 | Switzerland | 7 | 38 | Singapore | 2 |
| 15 | Belgium | 7 | 39 | Saudi Arabia | 2 |
| 16 | Portugal | 6 | 40 | Russian Federation | 2 |
| 17 | Hong Kong | 6 | 41 | Mexico | 2 |
| 18 | Thailand | 6 | 42 | Costa Rica | 1 |
| 19 | Poland | 5 | 43 | Colombia | 1 |
| 20 | France | 5 | 44 | Chile | 1 |
| 21 | Republic Of Korea | 5 | 45 | Slovenia | 1 |
| 22 | Denmark | 5 | 46 | Iceland | 1 |
| 23 | Austria | 5 | 47 | India | 1 |
| 24 | Finland | 5 | 48 | Croatia (local name: Hrvatska) | 1 |

Table 4: Average rank of countries in four editions in WR (12000 universities)

| | AR (WR,12000) | | | | | | | | | |
|---|---|---|---|---|---|---|---|---|---|---|
| Rank | Country Name | January 2012 | July 2012 | January 2013 | July 2013 | Rank | Country Name | January 2012 | July 2012 | January 2013 | July 2013 |
| 1 | Sweden | 3487.88 | 3549 | 3400.60 | 3777.52 | 26 | Belgium | 6012.53 | 6861.61 | 6137.20 | 6830.57 |
| 2 | Taiwan | 3543.68 | 3863.19 | 3668.77 | 4229.68 | 27 | Russian Federation | 6087.01 | 6039.20 | 6584.17 | 7496.95 |
| 3 | United Kingdom | 3954.39 | 4178.92 | 4632.36 | 4808.39 | 28 | Ireland | 6126.20 | 7253.33 | 5610.87 | 5563.92 |
| 4 | Finland | 4375.17 | 4721.65 | 4368.21 | 5734.67 | 29 | Venezuela | 6128.08 | 6457.65 | 6335.03 | 7240.48 |
| 5 | Turkey | 4402.29 | 4604.19 | 4891.14 | 5438.31 | 30 | Bulgaria | 6145.48 | 6277.72 | 5966.25 | 7265.13 |
| 6 | Australia | 4626.54 | 5116.29 | 4472.39 | 4634.24 | 31 | Japan | 6221.98 | 6681.09 | 6722.61 | 6865.12 |
| 7 | Greece | 4757.35 | 4840.66 | 4466.47 | 4829.35 | 32 | Indonesia | 6368.83 | 6228.70 | 6030.03 | 6415.21 |
| 8 | Italy | 4840.81 | 5463.68 | 4910.41 | 4902.27 | 33 | Netherlands | 6393.04 | 6768.57 | 5718.88 | 5881.94 |
| 9 | Norway | 4900.67 | 5030.90 | 4839.61 | 5138.66 | 34 | Ecuador | 6402.30 | 6648.16 | 5075.73 | 7041.13 |
| 10 | Thailand | 4938.87 | 5214.03 | 5594.10 | 5126.14 | 35 | Colombia | 6442.94 | 6753.41 | 6437.93 | 6822.69 |
| 11 | Germany | 5127.90 | 5448.34 | 5275.80 | 5460.31 | 36 | Switzerland | 6560.88 | 6946.32 | 6219.60 | 7058.88 |
| 12 | Canada | 5143.56 | 5067.65 | 4795.67 | 5817.54 | 37 | Vietnam | 6680.28 | 6816.17 | 7153.72 | 7100.75 |
| 13 | China | 5265.69 | 5482.91 | 5015.54 | 3973.40 | 38 | Peru | 6991.29 | 7236.08 | 6891.98 | 7948.81 |
| 14 | Spain | 5422.91 | 5581.17 | 4931.34 | 4494.01 | 39 | Denmark | 7156.16 | 7948.31 | 7370.40 | 7319.73 |
| 15 | Austria | 5427.41 | 5581.36 | 5481.92 | 5979.34 | 40 | Brazil | 7167.46 | 7487.41 | 6682.19 | 6465.13 |
| 16 | United States of America | 5429.12 | 5787.21 | 5687.48 | 5437.92 | 41 | Iran (Islamic Republic of Iran) | 7258.99 | 7651.30 | 6551.68 | 6976.69 |
| 17 | Malaysia | 5534.20 | 5111.09 | 4757.03 | 5593.72 | 42 | Poland | 7298.72 | 7322.05 | 6815.13 | 6995.31 |
| 18 | Chile | 5588.94 | 5551.67 | 4987.02 | 5194.88 | 43 | Mexico | 7328.01 | 7613.70 | 6499.49 | 6584.10 |
| 19 | Portugal | 5611.13 | 5746.15 | 5183.54 | 5418.44 | 44 | Republic Of Korea | 7349.12 | 6911.81 | 7183.57 | 6932.26 |
| 20 | Romania | 5804.34 | 6290.84 | 5799.40 | 6442.63 | 45 | Ukraine | 7381.49 | 6709.87 | 7249.29 | 7984.90 |
| 21 | Argentina | 5851.56 | 5686.82 | 5477.96 | 6267.42 | 46 | Philippines | 7863.76 | 7940.08 | 8369.33 | 7946.90 |
| 22 | France | 5857.22 | 6066.24 | 6192.60 | 6272.95 | 47 | India | 7964.18 | 8577.83 | 7746.40 | 7443.64 |
| 23 | Czech Republic | 5902.52 | 6177.15 | 5190 | 6556.94 | 48 | Pakistan | 8025.65 | 8549.49 | 7765.70 | 8021.74 |
| 24 | Hungary | 5993.70 | 6469.11 | 5949.71 | 6388.10 | 49 | Kazakstan | 8205.52 | 8323.27 | 7266.70 | 8489.78 |
| 25 | Egypt | 5998.91 | 6458.78 | 6697.35 | 5695.26 | 50 | Morocco | 9111.64 | 9365.50 | 8265.86 | 8413.03 |

Table 5: Weight of 50 countries in four editions in WR (12000 universities)

| Rank | Country Name | January 2012 | July 2012 | January 2013 | July 2013 | Rank | Country Name | January 2012 | July 2012 | January 2013 | July 2013 |
|---|---|---|---|---|---|---|---|---|---|---|---|
| 1 | United States of America | 18943854 | 17873913 | 17858132 | 19889656 | 26 | Netherlands | 504626 | 463593 | 489927 | 434382 |
| 2 | China | 7502026 | 7137381 | 7801639 | 8845309 | 27 | Switzerland | 456886 | 421019 | 404628 | 400231 |
| 3 | Japan | 3923274 | 3659314 | 4000265 | 3969265 | 28 | Vietnam | 452176 | 430748 | 402241 | 377242 |
| 4 | Brazil | 2783541 | 2702780 | 2733355 | 2025762 | 29 | Portugal | 447221 | 454617 | 470336 | 421220 |
| 5 | Russian Federation | 2643107 | 2635135 | 3466130 | 2598259 | 30 | Austria | 433791 | 425087 | 417157 | 385322 |
| 6 | Germany | 2315898 | 2212635 | 2178642 | 2230034 | 31 | Malaysia | 413811 | 441737 | 434578 | 410002 |
| 7 | France | 2285116 | 2229886 | 2305538 | 2256456 | 32 | Chile | 410308 | 408536 | 392727 | 387892 |
| 8 | United Kingdom | 1753944 | 1648411 | 1790337 | 1848245 | 33 | Belgium | 383198 | 335192 | 357631 | 299827 |
| 9 | India | 1525541 | 1401649 | 1624877 | 2177938 | 34 | Czech Republic | 371946 | 367462 | 347310 | 293925 |
| 10 | Republic Of Korea | 1330151 | 1469614 | 1310069 | 1241596 | 35 | Greece | 369375 | 359370 | 369143 | 372874 |
| 11 | Taiwan | 1302273 | 1253383 | 1258016 | 1196629 | 36 | Sweden | 366021 | 364288 | 369774 | 345344 |
| 12 | Poland | 1245839 | 1256980 | 1415469 | 1351266 | 37 | Finland | 365992 | 350003 | 358694 | 281940 |
| 13 | Canada | 1055891 | 1078127 | 1231940 | 1428148 | 38 | Norway | 362066 | 357531 | 350859 | 322483 |
| 14 | Spain | 1012872 | 978637 | 968406 | 983285 | 39 | Hungary | 342359 | 313578 | 314615 | 269371 |
| 15 | Italy | 1009446 | 938083 | 893288 | 936900 | 40 | Pakistan | 341794 | 329826 | 313338 | 338152 |
| 16 | Turkey | 995300 | 984947 | 995241 | 918637 | 41 | Peru | 315549 | 310029 | 326913 | 218764 |
| 17 | Thailand | 988558 | 960506 | 864796 | 893602 | 42 | Philippines | 297809 | 303569 | 232363 | 247239 |
| 18 | Mexico | 939070 | 923687 | 841578 | 758226 | 43 | Denmark | 271255 | 241579 | 217591 | 210612 |
| 19 | Indonesia | 912249 | 979807 | 1122355 | 893567 | 44 | Ecuador | 246299 | 231512 | 256198 | 148766 |
| 20 | Iran (Islamic Republic of Iran) | 843900 | 828029 | 877180 | 864010 | 45 | Ireland | 240826 | 206648 | 198063 | 231699 |
| 21 | Colombia | 772432 | 763760 | 739755 | 491844 | 46 | Bulgaria | 234181 | 246661 | 241350 | 189395 |
| 22 | Argentina | 602547 | 630319 | 626116 | 561793 | 47 | Venezuela | 217261 | 208483 | 181279 | 138026 |
| 23 | Australia | 530889 | 503528 | 526933 | 574529 | 48 | Egypt | 204037 | 199981 | 196198 | 214361 |
| 24 | Romania | 526631 | 494599 | 483647 | 372344 | 49 | Kazakstan | 166957 | 166902 | 189332 | 143919 |
| 25 | Ukraine | 517273 | 600858 | 874130 | 505902 | 50 | Morocco | 103981 | 105541 | 108290 | 129131 |

Table 6: Rank of 50 countries in four editions in WR (12000 universities)

| Country Name | January 2012 | July 2012 | January 2013 | July 2013 | Country Name | January 2012 | July 2012 | January 2013 | July 2013 |
|---|---|---|---|---|---|---|---|---|---|
| United States of America | 1 | 1 | 1 | 1 | Netherlands | 26 | 26 | 25 | 25 |
| China | 2 | 2 | 2 | 2 | Switzerland | 27 | 31 | 30 | 28 |
| Japan | 3 | 3 | 3 | 3 | Vietnam | 28 | 29 | 31 | 31 |
| Brazil | 4 | 4 | 5 | 8 | Portugal | 29 | 27 | 27 | 26 |
| Russian Federation | 5 | 5 | 4 | 4 | Austria | 30 | 30 | 29 | 30 |
| Germany | 6 | 7 | 7 | 6 | Malaysia | 31 | 28 | 28 | 27 |
| France | 7 | 6 | 6 | 5 | Chile | 32 | 32 | 32 | 29 |
| United Kingdom | 8 | 8 | 8 | 9 | Belgium | 33 | 38 | 36 | 37 |
| India | 9 | 10 | 9 | 7 | Czech Republic | 34 | 33 | 38 | 38 |
| Republic Of Korea | 10 | 9 | 11 | 12 | Greece | 35 | 35 | 34 | 32 |
| Taiwan | 11 | 12 | 12 | 13 | Sweden | 36 | 34 | 33 | 34 |
| Poland | 12 | 11 | 10 | 11 | inland | 37 | 37 | 35 | 39 |
| Canada | 13 | 13 | 13 | 10 | Norway | 38 | 36 | 37 | 36 |
| Spain | 14 | 16 | 16 | 14 | Hungary | 39 | 40 | 40 | 40 |
| Italy | 15 | 18 | 17 | 15 | Pakistan | 40 | 39 | 41 | 35 |
| Turkey | 16 | 14 | 15 | 16 | Peru | 41 | 41 | 39 | 43 |
| Thailand | 17 | 17 | 20 | 17 | Philippines | 42 | 42 | 44 | 41 |
| Mexico | 18 | 19 | 21 | 20 | Denmark | 43 | 44 | 45 | 45 |
| Indonesia | 19 | 15 | 14 | 18 | Ecuador | 44 | 45 | 42 | 47 |
| Iran (Islamic Republic of Iran) | 20 | 20 | 18 | 19 | Ireland | 45 | 47 | 46 | 42 |
| Colombia | 21 | 21 | 22 | 24 | Bulgaria | 46 | 43 | 43 | 46 |
| Argentina | 22 | 22 | 23 | 22 | Venezuela | 47 | 46 | 49 | 49 |
| Australia | 23 | 24 | 24 | 21 | Egypt | 48 | 48 | 47 | 44 |
| Romania | 24 | 25 | 26 | 33 | Kazakstan | 49 | 49 | 48 | 48 |
| Ukraine | 25 | 23 | 19 | 23 | Morocco | 50 | 50 | 50 | 50 |

## Comparing ranks of countries in top 500 WR and top 500 QS

Most of the researchers who analyze rankings of universities, use the data from the top 500 universities (http://webometrics.info/en/node/54, 2014). For the two rankings, WR and QS, we calculated the ranks rankings of countries based on the top 500 universities list. Table (7) shows the ranks rankings of countries in QS (2012) and table (8) shows the ranks rankings of countries in WR (January 2012).

M=500

Max W= M*(M-1)/2 =124750

Tables (7) and (8) show count the number of universities, weight of countries and total country's ranking in the top 500 WR and top 500 QS lists. The result shows that the rankings of two universities' rankings are highly similar.

Table 7: Rank of 50 countries (Top 500 universities of QS, 2012)

| Rank | Country Name | University | weight | Rank | Country Name | University | weight |
|---|---|---|---|---|---|---|---|
| 1 | United States of America | 99 | 28147 | 26 | Palestine | 4 | 1054 |
| 2 | United Kingdom | 51 | 15817 | 27 | Austria | 5 | 942 |
| 3 | Germany | 39 | 9417 | 28 | Singapore | 2 | 928 |
| 4 | Australia | 24 | 6193 | 29 | Brazil | 5 | 900 |
| 5 | Japan | 20 | 5385 | 30 | Saudi Arabia | 4 | 836 |
| 6 | Canada | 20 | 5307 | 31 | Argentina | 5 | 739 |
| 7 | France | 22 | 5001 | 32 | Chile | 3 | 605 |
| 8 | Netherlands | 13 | 4579 | 33 | South Africa | 3 | 558 |
| 9 | China | 18 | 3431 | 34 | Mexico | 2 | 548 |
| 10 | Republic Of Korea | 13 | 3370 | 35 | Thailand | 2 | 544 |
| 11 | Switzerland | 8 | 3016 | 36 | Colombia | 4 | 383 |
| 12 | Sweden | 8 | 2482 | 37 | Indonesia | 3 | 327 |
| 13 | Hong Kong | 6 | 2379 | 38 | Portugal | 3 | 265 |
| 14 | Belgium | 7 | 2347 | 39 | Lebanon | 1 | 250 |
| 15 | Spain | 14 | 2087 | 40 | Kazakstan | 2 | 241 |
| 16 | Italy | 14 | 2017 | 41 | Czech Republic | 1 | 214 |
| 17 | Taiwan | 11 | 1864 | 42 | United Arab Emirates | 2 | 205 |
| 18 | Finland | 8 | 1796 | 43 | Philippines | 2 | 177 |
| 19 | New Zealand | 7 | 1669 | 44 | Poland | 2 | 177 |
| 20 | Denmark | 5 | 1557 | 45 | Egypt | 1 | 108 |
| 21 | Ireland | 7 | 1549 | 46 | Turkey | 2 | 100 |

| Rank | Country Name | University | weight | Rank | Country Name | University | weight |
|---|---|---|---|---|---|---|---|
| 22 | India | 7 | 1272 | 47 | Oman | 1 | 75 |
| 23 | Norway | 4 | 1151 | 48 | Pakistan | 1 | 75 |
| 24 | Malaysia | 6 | 1114 | 49 | Greece | 2 | 50 |
| 25 | Russian Federation | 6 | 1066 | 50 | Uruguay | 1 | 25 |

Table 8: Rank of 48 countries (Top 500 universities of WR, January 2012)

| Rank | Country Name | University | weight | Rank | Country Name | University | weight |
|---|---|---|---|---|---|---|---|
| 1 | United States of America | 155 | 47285 | 25 | Finland | 5 | 974 |
| 2 | Germany | 44 | 9874 | 26 | Poland | 5 | 824 |
| 3 | United Kingdom | 30 | 6906 | 27 | Greece | 4 | 795 |
| 4 | Canada | 23 | 6852 | 28 | Indonesia | 3 | 609 |
| 5 | Spain | 24 | 5135 | 29 | Singapore | 2 | 522 |
| 6 | Australia | 17 | 3837 | 30 | New Zealand | 3 | 512 |
| 7 | Taiwan | 14 | 3643 | 31 | Mexico | 2 | 469 |
| 8 | Italy | 16 | 3425 | 32 | Ireland | 4 | 461 |
| 9 | Brazil | 12 | 3403 | 33 | Hungary | 3 | 426 |
| 10 | Netherlands | 11 | 3113 | 34 | Slovenia | 1 | 420 |
| 11 | China | 14 | 2740 | 35 | France | 5 | 390 |
| 12 | Japan | 12 | 2690 | 36 | Russian Federation | 2 | 388 |
| 13 | Sweden | 9 | 2425 | 37 | Chile | 1 | 347 |
| 14 | Switzerland | 7 | 1805 | 38 | Saudi Arabia | 2 | 328 |
| 15 | Thailand | 6 | 1788 | 39 | Turkey | 4 | 265 |
| 16 | Belgium | 7 | 1659 | 40 | Argentina | 2 | 250 |
| 17 | Hong Kong | 6 | 1444 | 41 | South Africa | 3 | 185 |
| 18 | Portugal | 6 | 1371 | 42 | Malaysia | 3 | 171 |
| 19 | Austria | 5 | 1348 | 43 | Colombia | 1 | 143 |
| 20 | Denmark | 5 | 1155 | 44 | Costa Rica | 1 | 97 |
| 21 | Norway | 4 | 1144 | 45 | Croatia (local name: Hrvatska) | 1 | 54 |
| 22 | Czech Republic | 3 | 1065 | 46 | India | 1 | 46 |

| Rank | Country Name | University | weight | Rank | Country Name | University | weight |
|---|---|---|---|---|---|---|---|
| 23 | Palestine | 4 | 1029 | 47 | Slovakia | 2 | 42 |
| 24 | Republic Of Korea | 5 | 984 | 48 | Iceland | 1 | 19 |

## Limitations and Discussion

Changing methodology of the rankings of universities are the study's main limitations. Results which have been calculated based on each edition are not comparable. For comparing rankings of universities in different editions we should re-rank previous data with the new methodology. Characteristics of the Web are second limitation. We know search engines' results are not stable and changing continuously. Also, data sources' policies change among the time and providers lead to use new data sources for attracting web statistics. Rankings of universities which are based on web data sources only valid for the time span which they are collected and result calculated.

## Conclusion

In this paper, we introduced the new rankings of countries methods. Average rank of universities (AR) and weighting countries (W) based on their university ranks. The average rank is simple but not suitable to show a good view of countries rank. Using W in one ranking of universities in different editions shows us the result of the activities of countries to take better rank. The current study covered two universities' rankings which are very different methodologies. We compared different world universities' rankings (QS & WR) using W. The results show that the rankings of two rankings lists (WR and QS) are highly similar. On the other hand, deposit each ranking used different methods. This gives some confidence that to W is a robust algorithm to rank all countries around the world.